\newcommand\ezpm{E_0^{\pm}(r)}
\newcommand\ezp{E_0^{+}(r)}
\newcommand\ezm{E_0^{-}(r)}
\newcommand\ezph{E_0^{+}(r_+)}
\newcommand\ezmh{E_0^{-}(r_+)}

\newcommand\rp{r_+}
\newcommand\rmen{r_{-}}

\newcommand\pa{\partial}

\newcommand\beque{\begin{equation*}}
\newcommand\beq{\begin{equation}}
\newcommand\eeq{\end{equation}}
\newcommand\eeque{\end{equation*}}
\newcommand\beqnl{\begin{eqnarray}}
\newcommand\beqna{\begin{eqnarray*}}
\newcommand\eeqna{\end{eqnarray*}}
\newcommand\eeqnl{\end{eqnarray}}

 \def\NN{\hbox{\sf I\kern-.13em\hbox{N}}}
 \def\HH{\hbox{\sf I\kern-.13em\hbox{H}}}
 \def\DD{\hbox{\sf I\kern-.13em\hbox{D}}}
 \def\RR{\hbox{\sf I\kern-.14em\hbox{R}}}
 \def\CC{\hbox{\sf I\kern-.44em\hbox{C}}}
 \def\ZZ{{\hbox{\sf Z\kern-.43emZ}}}
 \def\QQ{\hbox{\sf C\kern -.48emQ}}
 \def\Cc{\hbox{\sf C\kern -.47em {\raise .48ex \hbox{$\scriptscriptstyle |$}}
   \kern-.5em {\raise .48ex \hbox{$\scriptscriptstyle |$}} }}
 \def\Qq{\hbox{\sf Q\kern -.57em {\raise .48ex \hbox{$\scriptscriptstyle |$}}
   \kern-.55em {\raise .48ex \hbox{$\scriptscriptstyle |$}} }}

\documentclass[12pt]{iopart}
\usepackage{amsfonts}
\usepackage{setstack}
\usepackage{amsthm}
\usepackage{amsbsy}
\usepackage{amssymb}
\usepackage{epsfig}

\begin{document}

\date{\today}

\title[Quantum Dirac Field  in Reissner-Nordstr\"{o}m-AdS Black 
Hole Background]{Quantum Effects for the Dirac Field  in Reissner-Nordstr\"{o}m-AdS Black 
Hole Background}

\author{F. Belgiorno\footnote{E-mail address: belgiorno@mi.infn.it}}
\address{Dipartimento di Fisica, Universit\`a di Milano, 20133 
Milano, Italy, and\\
I.N.F.N., sezione di Milano, Italy}
\author{S. L. Cacciatori\footnote{E-mail address: sergio.cacciatori@uninsubria.it}}
\address{Dipartimento di Fisica, Universit\`a dell'Insubria, 22100  
Como, Italy, and\\
I.N.F.N., sezione di Milano, Italy}

\begin{abstract}
The behavior of a charged massive Dirac field  
on a Reissner-Nordstr\"{o}m-AdS black hole background 
is investigated. The essential self-adjointness of the Dirac 
Hamiltonian is studied. Then, an analysis of the discharge problem 
is carried out in analogy with the standard 
Reissner-Nordstr\"{o}m black hole case. 

\end{abstract}

\maketitle

\section{Introduction}
\label{intro}

AdS background geometry has been considering a challenging field for quantum field theory 
in different frameworks, including supersymmetry and string theory. In this paper, 
we take into account the problem of the discharge of a Reissner-Nordstr\"{o}m-AdS black hole 
by quantum effects, by extending the analysis carried by \cite{gibbons,khriplovich} 
to the case of a charged black hole which is asymptotically AdS. This kind of analysis, which 
is addressed to the case of massive charged Dirac fields,  
represents also a completion on the quantum side of the analysis carried out in 
Ref. \cite{winstanley}, where classical superradiance effects in a Kerr-Newman-AdS black hole 
background are considered for the case of an uncharged scalar field, both in the case of reflecting 
boundary conditions and in the case of transparent boundary conditions. 
Herein, the behavior of a charged massive Dirac field  
on a Reissner-Nordstr\"{o}m-AdS black hole background 
is investigated. The essential self-adjointness of the Dirac 
Hamiltonian is studied. Then, an analysis of the discharge problem 
is carried out in analogy with the standard 
Reissner-Nordstr\"{o}m black hole case. 

\section{Dirac Hamiltonian}
\label{essauto}

In this section we check if the one-particle Hamiltonian 
is well-defined in the sense that no boundary conditions 
are required in order to obtain a self-adjoint operator. 
In other terms, we check if the 
Hamiltonian is essentially self-adjoint, that is,  
if a unique self-adjoint extension and a uniquely determined 
physics occur. A general study for Dirac equation on the Kerr-Newman-AdS black hole 
background is progress \cite{cacciatori}.  
\\ 
We first define the one-particle Hamiltonian for 
Dirac massive particles on the Reissner-Nordstr\"{o}m-CAdS black hole geometry, to 
which we refer for simplicity as RN-AdS in the following. 
We use natural units ${\mathrm \hbar}={\mathrm c}={\mathrm G}=1$ 
and unrationalized electric units. 
The metric of the RN-AdS manifold 
$(t\in R;\ r\in (r_+,+\infty);\ \Omega\in S^2)$ is 
\beqnl
ds^2&=& -f(r) dt^2+\frac{1}{f(r)} dr^2 +r^2 d\Omega^2\cr
f(r)&=& 1+\frac{r^2}{L^2}-\frac{2 M}{r}+\frac{Q^2}{r^2};
\label{regeo}
\eeqnl
$r_+$ is the radius of the black hole event horizon, 
$L$ is a length associated with the cosmological constant $\Lambda$ by 
$\Lambda=-3/L^2\Leftrightarrow L=\sqrt{\frac{3}{|\Lambda|}}$; $M$ is the mass and $Q$ is the 
electric charge. 
The vector potential associated with the 
RN-AdS solution is $A_{\mu}=(-Q/r,0,0,0)$. 
The spherical symmetry of the problem allows to 
separate the variables 
and to study a reduced problem on a fixed eigenvalue sector of the 
angular momentum operator. For a complete deduction of the 
variable separation see e.g. 
\cite{soffel,thaller}. We get the following reduced Hamiltonian 
\[ H_{red}=\left[
\begin{array}{cc}
\sqrt{f}\; \mu+ \frac{e\; Q}{r} &\   
-f\; \partial_r +k\; \frac{\sqrt{f}}{r}\cr
f\; \partial_r +k\; \frac{\sqrt{f}}{r}  
&\  -\sqrt{f}\; \mu+\frac{e\; Q}{r}
\end{array} 
\right] 
\]
where  $f(r)$ is the same as in (\ref{regeo}), 
$k=\pm (j+1/2) \in \ZZ-\{0\}$ is the 
angular momentum eigenvalue and $\mu$ is the mass of the Dirac particle.  The Hilbert space in which 
$H_{red}$ is formally defined 
is the Hilbert space $L^2 [(r_+,+\infty), 1/f(r)\; dr]^{2}$ 
of the two-dimensional 
vector functions $\vec{g}\equiv \left(\begin{array}{c}
g_1 \cr
g_2 \end{array}\right)$ such that 
$$
\int_{r_+}^{+\infty}\; \frac{dr}{f(r)}\; (|g_1 (r)|^2+|g_2 (r)|^2)<\infty.
$$
As a domain for the minimal operator associated with  
$H_{red}$ we can choose the following subset 
of $L^2 [(r_+,+\infty), 1/f(r)\; dr]^{2}$: the set    
$C_{0}^{\infty}(r_+,+\infty)^2$ 
of the two-dimensional vector functions 
$\vec{g}$ whose components are smooth and of compact support 
\cite{weidmann}. 
It is useful to define a new tortoise-like variable $y$ 
\beq
\frac{dy}{dr}=-\frac{1}{f(r)}
\eeq
and to choose an arbitrary integration constant in such a way 
that $y\in (0,+\infty)$. The reduced Hamiltonian becomes 
\beq
H_{red}=D_{0}+V(y)
\label{red}
\eeq
where
\[ D_{0}=\left[
\begin{array}{cc}
0 &\   \partial_y \cr 
-\partial_y &\ 0\end{array} \right] 
\]
and 
\[ V(r(y))=\left[
\begin{array}{cc}
\sqrt{f}\; \mu+ \frac{e\; Q}{r} &\  k\; \frac{\sqrt{f}}{r} \cr 
k\; \frac{\sqrt{f}}{r}
&\ -\sqrt{f}\; \mu+\frac{e\; Q}{r}\end{array} \right]. 
\]
The Hilbert space of interest for the 
Hamiltonian (\ref{red}) is $L^2 [(0,+\infty), dy]^{2}$. 
We have to check if the reduced Hamiltonian is essentially self-adjoint;  
with this aim, we check if the solutions of the equation
\beq
H_{red}\; g=\lambda\; g
\label{eigen}
\eeq
are square integrable in a right neighborhood of $y=0$ and in 
a left neighborhood of $y=+\infty$. 
The so called Weyl alternative generalized to a system of first order 
ordinary differential equations (\cite{weidmann}, theorem 5.6) 
states that the so-called limit circle case (LCC) occurs 
at $y=0$ if for every $\lambda\in \CC$ all the solutions of 
$(H_{red}-\lambda) g=0$ lie in $L^2[(0,b), dy]^{2}$ in a right neighborhood $(0,b)$ of  
$y=0$. If at least one solution not square 
integrable exists for every $\lambda \in \CC$, then no 
boundary condition is required and 
the so-called limit point case (LPC) is verified. 
Note that, if for a fixed $\lambda_0 \in \CC$ all the solutions of 
$(H_{red}-\lambda_0) g=0$ and of $(H_{red}-\bar{\lambda}_0) g=0$
lie in $L^2[(0,b), dy]^{2}$ in a right neighborhood $(0,b)$ of  
$y=0$, then this holds true for any $\lambda\in \CC$ (\cite{weidmann}, theorem 5.3).  
The occurrence of LCC implies the necessity to 
introduce boundary conditions in order to 
obtain a self-adjoint operator. 
If at least one solution not square 
integrable exists for every $\lambda \in \CC$, then no 
boundary condition is required and 
the so-called limit point case (LPC) is verified. The 
same arguments  can be applied for $y=+\infty$. 
The Hamiltonian operator is essentially self-adjoint if the LPC 
is verified both at $y=0$ and at infinity 
(cf. \cite{weidmann}, theorem 5.7).

\subsection{RN-AdS: behavior at $r=\infty$}
\label{sect-infinity}

In order to determine the essential self-adjointness properties of the Dirac operator in 
the RN-AdS black hole background, a study of the behavior of the first order differential system 
(\ref{eigen}) near $r=\infty$ has to be done. The substitution $x=1/r$ allows to 
map the interval $(r_+,\infty)$ into the interval $(0,1/r_+)$. The reduced 
Hamiltonian is formally self-adjoint in $L^2 ((0,1/r_+), 1/(x^2\; f(x))\; dx)^2$. Then 
the behavior as $x\to 0$ of the solutions has to be determined.  
In particular the eigenvalue equation can be re-written as follows:
\[ x \vec{g}'=\left[
\begin{array}{cc}
\frac{k L x}{\sqrt{h(x)}} &\ 
-\frac{1}{\sqrt{h(x)}}\; \mu L+ \frac{e\; Q L^2 x^2}{h(x)} -\lambda\; \frac{L^2 x}{h(x)} \cr 
-\frac{1}{\sqrt{h(x)}}\; \mu L- \frac{e\; Q L^2 x^2}{h(x)} +\lambda\; \frac{L^2 x}{h(x)} & 
-\frac{k L x}{\sqrt{h(x)}}
\end{array} \right] \vec{g},  
\]
where the prime indicates a derivative with respect to $x$ and $h(x)$ is 
defined by 
\beq
h(x) = 1+L^2 x^2 - 2 M L^2 x^3 +Q^2 L^2 x^4 = L^2 x^2 f(x). 
\eeq
The matrix on the left of the above system is regular at $x=0$ and its limit as $x\to 0$ 
is given by the constant matrix 
\[ A_0=\left[
\begin{array}{cc}
0  &
-\mu L\cr 
-\mu L & 0 
\end{array} \right].   
\]
The given linear system displays a singularity of the first kind at $x=0$ \cite{hsieh}, also 
called weakly singular point \cite{walter}. 
Then the eigenvalues $\epsilon_{\pm} = \pm \mu L$ of the matrix $A_0$  determine the 
asymptotic behavior of the solutions of the system near $x=0$ (cf. thm. V-1-2 p.110 of Ref. 
\cite{hsieh}), which, near $x=0$ behave as follows: if $\epsilon_{+}-\epsilon_{-}=2 \mu L \ne 
m$, with $m \in \NN-\{0\}$, then two linearly independent solutions are of the form 
\beqnl
\vec{g}_{(+)} (x) &=& x^{\epsilon_{+}} \vec{w}(x)\\
\vec{g}_{(-)} (x) &=& x^{\epsilon_{-}} \vec{v}(x),
\eeqnl
where $\vec{w}(x)=\sum_{k=0}^{\infty} \vec{w}_k x^k,\vec{v}(x)=\sum_{k=0}^{\infty}\vec{v}_k x^k$ are 
(formal) series. If $\epsilon_{+}-\epsilon_{-}=m$, with $m\in \NN-\{0\}$, 
then two linearly independent solutions are of the form 
\beqnl
\vec{g}_{(+)} (x) &=& x^{\epsilon_{+}} \vec{w}(x)\\
\vec{g}_{(-)} (x) &=& x^{\epsilon_{-}} \vec{v}(x)+ c \vec{g}_{(+)} (x) \log (x).
\eeqnl
One finds the essential self-adjointness condition 
\beq
\mu L \geq \frac{1}{2}.
\eeq
It is worth pointing out that this condition coincides with the one occurring in the 
pure AdS case (cf. \cite{bachelot} and see also Appendix A). 
Note that, if the condition $\mu>0$ 
on the mass is relaxed, then the Dirac Hamiltonian is self-adjoint for 
$|\mu L|\geq \frac{1}{2}$, as is it easy to show.  

\subsection{RN-AdS: behavior at $r=r_+$}

We introduce again the tortoise-like coordinate $y$. 
We find that 
\[ H_{red}=\left[
\begin{array}{cc}
\sqrt{f}\; \mu+ \frac{e\; Q}{r} &\  
+\partial_y +k\; \frac{\sqrt{f}}{r}\cr
-\partial_y +k\; \frac{\sqrt{f}}{r}  
&\ -\sqrt{f}\; \mu+\frac{e\; Q}{r}
\end{array} 
\right] 
\]
In order to study the behavior on the horizon, i.e. for $y\to \infty$, 
one can simply apply the corollary to thm. 6.8 (p. 99) of \cite{weidmann} 
and see that the LPC occurs on the horizon\footnote{Theorem 6.8 of \cite{weidmann} 
states that, given a Dirac system $\tau \vec{u}=B^{-1} \left[ J \vec{u}' + P \vec{u}\right]$, 
with $x\in (a,b)$ and $J=\left(
\begin{array}{cc}
0 &  
1\cr
-1  
& 0
\end{array} 
\right),$ 
if $B=k(x) \left(
\begin{array}{cc}
1 &  
0\cr
0  
& 1
\end{array} 
\right),$ 
with $k(x)\not \in L^1 (c,b)$ for all $c\in (a,b)$, then $\tau$ is in the LPC at $b$. 
As a corollary, if $b=\infty$ and $k(x)=d>0$, with $d=$const., then $\tau$ is in the LPC 
at $b=\infty$.}. This holds true both in the non-extremal case and in the extremal one. 
Thus, the Dirac operator 
is essentially self-adjoint on the RN-AdS black hole background if the 
bound on $\mu$ which is found in the pure AdS case is implemented. 
An alternative proof is found in Appendix B. 

\subsection{Essential spectrum from near the horizon}
\label{sub-essential}

One expects that, in presence of an event horizon, i.e. of a so-called ergosurface, 
the mass gap vanishes and that the continuous spectrum includes the whole real line. 
We recall that qualitative spectral methods for the Dirac equation (see e.g. 
\cite{thaller,weidmann}) 
have been applied to Dirac fields 
on a black hole background in \cite{belgio,yamada}. 
In order to verify this property, we adopt the decomposition method. We introduce the 
restriction of the Hamiltonian near the horizon, say to the interval $(\rp,r_0)$, where 
$r_0$ is arbitrary, and near infinity, say to the interval $(r_0,\infty)$. For our 
purposes, it is sufficient to consider the former restriction. In the tortoise-like coordinate $y$ 
one finds a potential $P$ such that 
\[ P=\left[
\begin{array}{cc}
\sqrt{f}\; \mu+ \frac{e\; Q}{r} &\  
k\; \frac{\sqrt{f}}{r}\cr
k\; \frac{\sqrt{f}}{r}  
&\  -\sqrt{f}\; \mu+\frac{e\; Q}{r}
\end{array} 
\right] 
\]
and it holds 
\[ \lim_{y\to \infty} P(y) = P_0 =\left[
\begin{array}{cc}
\Phi_+  & 0\cr 
0 & \Phi_+
\end{array} 
\right] 
\]
which is in diagonal form and whose eigenvalues coincide. 
We apply theorem 16.6 p. 249 of Ref. \cite{weidmann}, which implies that, if 
$\nu_{-},\nu_+$, with $\nu_{-}\leq \nu_+$, are the eigenvalues of the matrix $P_0$, then 
$\{\RR-(\nu_{-},\nu_+)\} \subset \sigma_{e} (H_{red}|_{(y(r_0),\infty)})$ if 
\beq
\lim_{y\to \infty} \frac{1}{y} \int_{\epsilon_0}^y dt ||P(t)-P_0||=0,
\eeq
where $||\cdot||$ stays for any norm in the set of $2\times 2$ matrices (we choose the 
Euclidean norm). In our case 
one has to find the limit as $y\to \infty$ for the following expression:
\beq
\frac{1}{y} \int_{r(y)}^{r_1} dr \frac{1}{h(r)} \frac{1}{\sqrt{r-\rp}} \sqrt{2 \mu^2 h(r) +
2 \left( \Phi_+^2 (r-\rp)+k^2 h(r)\right) \frac{1}{r^2}},
\eeq
where we put $h(r)=\frac{f(r)}{r-\rp}$.  
In the non extremal case $\rp>\rmen$, the above integral is finite as $r\to \rp$ and then the limit 
is zero. In the extremal case $\rp=\rmen$ the integral above displays a logarithmic divergence 
as $r\to \rp$; a trivial use of the l'Hospital's rule allows to find that the aforementioned 
limit is still zero. As a consequence, we can state that 
\beq
\sigma_{e} (H_{red})=\RR.
\eeq

\section{Pair Creation and Level-Crossing}

In this section, we limit ourselves to the study of the case where essential 
self-adjointness is implemented. In particular, 
in the case of the Dirac field we impose 
$\mu L > 1/2$.\\ 
In the RN black hole background, the presence of an effect of 
particle creation can be related to the Klein paradox, i.e. to the 
possibility to find regions where negative energy states overlap 
with positive energy states.\\ 
We follow Ruffini-Damour-Deruelle approach \cite{chruffini,damo,deruelle,ruffini}, 
in which one identifies 
the effective potentials $E_0^{\pm}(r)$ for the positive and negative 
energy states respectively; they represent the classical turning points 
for the ptc. motion and lead to the definition of the so-called effective 
ergosphere. These potentials enter the Hamilton-Jacobi (HJ) equation for a 
classical particle. They can be interpreted also at the quantum level, as 
in \cite{ruffini}. In particular, they indicate the regions 
of level-crossing between positive and negative energy states 
\cite{damo,deruelle}. See also \cite{deru}. 
We give some details below.\\ 
As far as the Dirac equation is concerned, it is 
known that the HJ equation corresponds to a WKB approximation to the 
Dirac equation at the lowest order \cite{pauli,rubinow}. Cf. also \cite{maslov} 
and \cite{bolte} for a semiclassical approach to the Dirac equation. 
We first recall the HJ 
equation and find $E_0^{\pm}(r)$, then we show that it is possible 
to find a level-crossing even in the RN-AdS case. 
The HJ equation for a classical ptc. is 
\beq
g^{\mu \nu} (\pa_{\mu} S - e A_{\mu})(\pa_{\nu} S -e A_{\nu})+\mu^2=0.
\eeq
Variable separation leads to $S=-\omega t+\Theta (\theta,\phi)+R(z)$, 
where $dz=-dr/\Delta$, and
\beq
\left( \frac{dR}{dz} \right)^2=-Z,
\eeq
where 
\beq
Z\equiv \Delta (\mu^2 r^2+K)-[\omega r^2-e Q r]^2,
\eeq
and, in the RN case, $\Delta=r^2-2Mr+Q^2$  
($K$ is square of the angular momentum vector) \cite{damo}. 
The classical region of accessibility is defined by $Z\leq 0$. One has 
\beq
\Delta^2 \left( \frac{dR}{dr} \right)^2=r^4 (E-E_0^+ (r))(E-E_0^- (r)),
\eeq
where 
\beq
E_0^{\pm}(r)=\frac{eQ}{r}\pm \frac{1}{r^2} \sqrt{\Delta (\mu^2 r^2+K)}.
\eeq
From a classical point of view, $\ezpm$ represent effective potentials for 
the solutions having positive and negative energy, and particles states 
are defined for $E>\ezp$. Classical bound states correspond to circular 
or elliptical orbits and require the presence of a couple of turning points.  
States such that 
$\ezm<E<\ezp$ are forbidden (they correspond to particles having imaginary 
momentum). States with $E<\ezm$ are also classically forbidden, because 
they correspond to particles with negative mass (and negative energy). 
The latter states are meaningful at the quantum level (anti-particles)\cite{damo,deruelle}.\\ 
Variable separation in the quantum case allows to obtain an obvious improvement 
of the above classical formulas, amounting in replacing the classical value 
of $K$ with the quantum eigenvalues of the corresponding quantum operator $\hat{K}$. 
The analogy with the replacement of the classical centrifugal term 
$\frac{\vec{L}^2}{2 m r^2}$ with the quantum term $\frac{\hbar^2 l(l+1)}{2 m r^2}$ 
in the spherosymmetric case for the Schr\"{o}dinger equation is self-evident. 
In subsection \ref{meaning}, the meaning of the aforementioned improved potentials 
will be discussed. 

It can be shown that it is possible to find level-crossing in the RN-AdS 
case, where $\Delta=r^2-2Mr+Q^2+r^4/L^2$. In particular, 
a level crossing surely occurs when $E_0^+ (r)<\ezph =eQ/r_+$ 
for some $r>r_+$. See below. Notice also that $\ezmh =eQ/r_+$, 
i.e., both the positive and negative energy states assume the same 
value on the horizon (this corresponds to the well-known spectral 
property that the contribution to the essential spectrum arising from 
near the horizon is $\RR$).\\

\subsection{Level crossing}

It is useful to define for $e Q>0$
\beq
P(r)\equiv \ezp-\frac{eQ}{r_+}=-\frac{eQ}{r r_+} (r-\rp)+
\frac{1}{r^2} \sqrt{\Delta (\mu^2 r^2+K)},
\eeq
and 
\beq
N(r)\equiv \ezm-\frac{eQ}{r_+}=-\frac{eQ}{r r_+} (r-\rp)-
\frac{1}{r^2} \sqrt{\Delta (\mu^2 r^2+K)}.
\eeq
It is evident that, for any $r>r_+$, one has $N(r)<0$ (for $e Q<0$ an obvious 
change in the definitions above is required). 
This means that 
$P(r)<0$ is the condition which has to be satisfied in order to find 
a level-crossing ($P(r)\geq 0$ for all $r>r_+$ means $E_0^+>E_0^-$ in the 
same region, i.e., no level-crossing). One has to study the roots of 
the equation $P(r)=0$ for $r\geq r_+$ and then find out the sign 
of $P(r)$ between two consecutive roots. Of course $r=r_+$ is a root 
of $P(r)$. It is necessary to find another root for $r>r_+$ in order to 
find level-crossing. 
It is useful to write
\beq
\Delta = (r-\rp) (r-r_{-}) G(r),
\eeq
where $G(r)\equiv (1/L^2) [r^2+(\rp+\rmen) r+L^2+\rp^2+\rmen^2+\rp \rmen]$. $\rp$ and $\rmen$ 
are the event horizon radius and the Cauchy horizon radius respectively. 
This re-writing takes into account that a term proportional to $r^3$ is missing in $\Delta$, and it 
amounts to the following re-parameterization of $f(r)$ in terms of $L^2,\rp,\rmen$:
\beq
f(r)=\frac{1}{L^2 r^2} (r-\rp) (r-r_{-}) [r^2+(\rp+\rmen) r+L^2+\rp^2+\rmen^2+\rp \rmen],
\eeq
with 
\beqnl
M\hphantom{'} &=& \frac{(\rp+\rmen)(\eta - \rp \rmen)}{2 L^2}\\
Q^2 &=& \frac{\eta \rp \rmen}{L^2},
\eeqnl
where 
\beq
\eta =  L^2+ \rp^2 + \rmen^2 + \rp \rmen.
\eeq
One has to implement $P(r)<0$, which amounts to
\beq
\frac{1}{r^2 L^2} (r-\rp) (r-\rmen) (r^2+(\rp+\rmen) r +\eta) (\mu^2 r^2+k^2) < 
\left( \frac{e Q}{\rp} \right)^2 (r-\rp)^2; 
\eeq
then, for the extremal case one obtains 
\beq
\frac{1}{r^2 L^2}  (r^2+2\rp r +3 \rp^2+L^2) (\mu^2 r^2+k^2) < 
\left( \frac{e Q}{\rp} \right)^2,  
\eeq
i.e.
\beqna
&&\mu^2 r^4 + 2 \rp \mu^2 r^3+
\left((3 \rp^2+L^2) \mu^2+k^2-\left( \frac{e Q}{\rp} \right)^2 L^2 \right) r^2 \cr
&&+2 k^2 \rp r 
+k^2 (3 \rp^2+L^2) <0.
\eeqna
From a naive inspection, one sees that the following necessary condition has to be implemented: 
\beq
(3 \rp^2+L^2) \mu^2+k^2-\left( \frac{e Q}{\rp} \right)^2 L^2 <0. 
\eeq
A more accurate study of the first derivative of $\ezp$ shows that 
\beq
\frac{dE^+_0 }{dr} (r=\rp) <0 
\eeq
is a sufficient condition for the existence of a level crossing, which 
amounts to 
\beq
\left( \frac{e Q}{\rp} \right)^2>\mu^2+\frac{6 \mu^2 \rp^2}{L^2}+\frac{6 k^2}{L^2}
+\frac{k^2}{\rp^2}.
\eeq
When this sufficient condition is implemented, the standard condition 
\beq
\left(\frac{e Q}{\rp}\right)^2 > \mu^2, 
\eeq
for the standard Reissner-Nordstr\"{o}m case is also implemented. 

\begin{figure}[h]
\setlength{\unitlength}{1.0mm}
\centerline{\psfig{figure=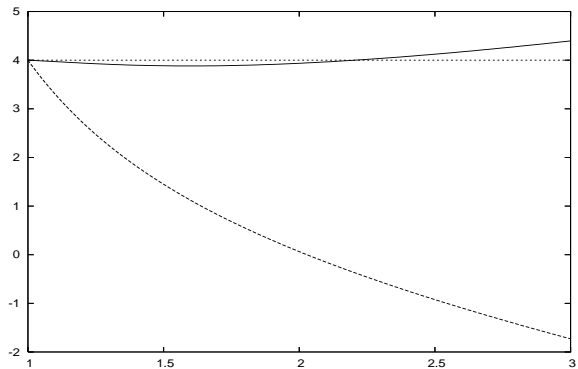,height=8cm,width=5cm,angle=-90}}
\vspace{0.2cm}
\caption{Level-crossing in the case of an extremal RN-AdS black hole, 
with $L=1,M=3,Q=2,m=1,e=2,k=1$. The particle and the 
black hole have charges with the same sign. Black hole horizon occurs for $r=1$.  
The straight line represents the value $e Q/r_+$. 
The upper potential is 
$\ezp$, the lower one is $\ezm$. Level-crossing occurs 
where $\ezp<e Q/r_+$.}
\label{fig1}
\end{figure}

\begin{figure}[h]
\setlength{\unitlength}{1.0mm}
\centerline{\psfig{figure=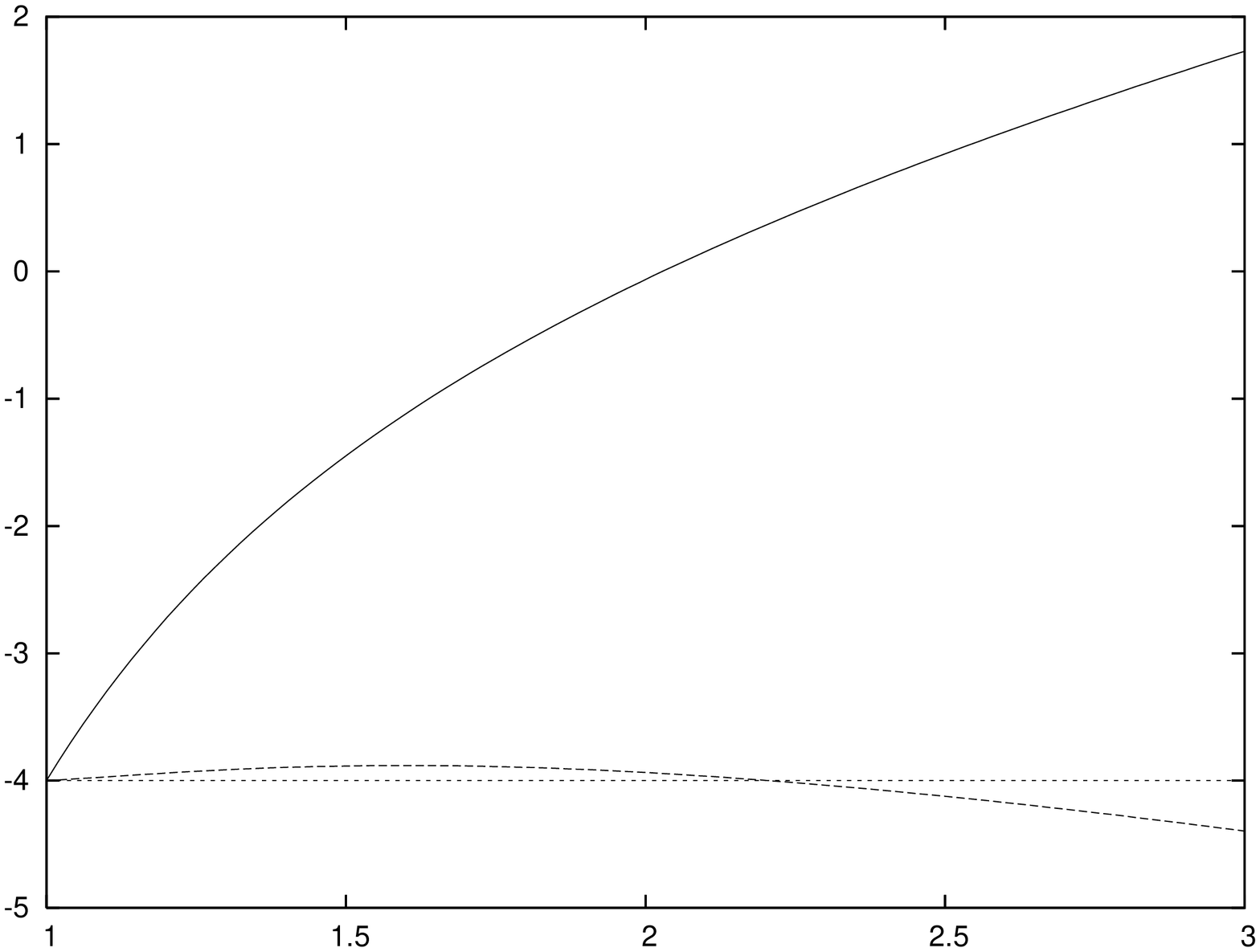,height=8cm,width=5cm,angle=-90}}
\vspace{0.2cm}
\caption{Level-crossing in the case of an extremal RN-AdS black hole, 
with $L=1,M=3,Q=2,m=1,e=-2,k=1$. The particle and the 
black hole have charges with opposite sign. Black hole horizon occurs for $r=1$.  
The straight line represents the value $e Q/r_+$. 
The upper potential is 
$\ezp$, the lower one is $\ezm$. Level-crossing occurs 
where $\ezp<e Q/r_+$.}
\label{fig2}
\end{figure}

\begin{figure}[htbp!]
\begin{picture}(150,150)
\put(0,160){\includegraphics[height=.34\textheight,
                      width=.24\textheight,angle=-90]{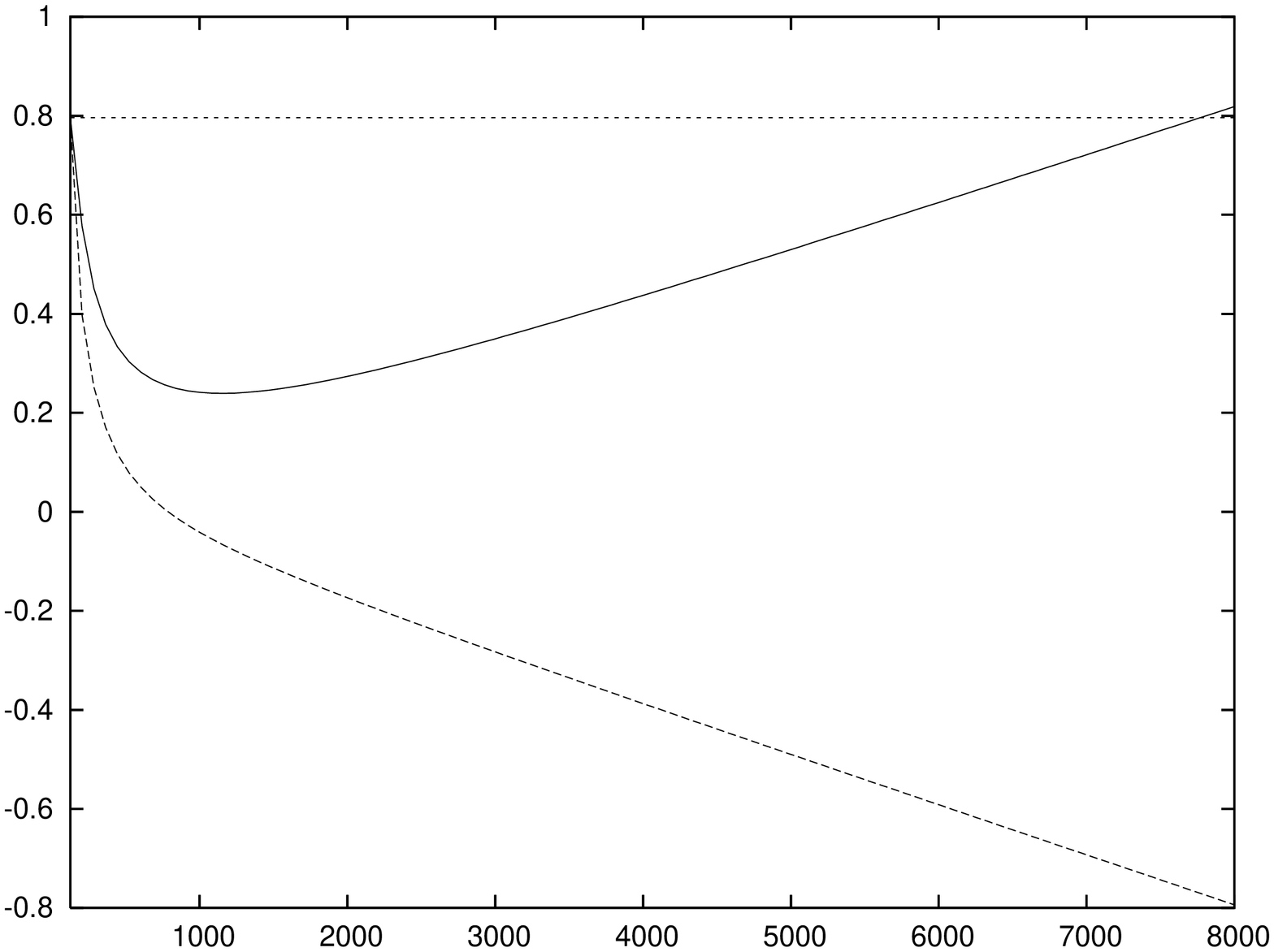}}
\put(230,160){\includegraphics[height=.30\textheight,
                      width=.24\textheight,angle=-90]{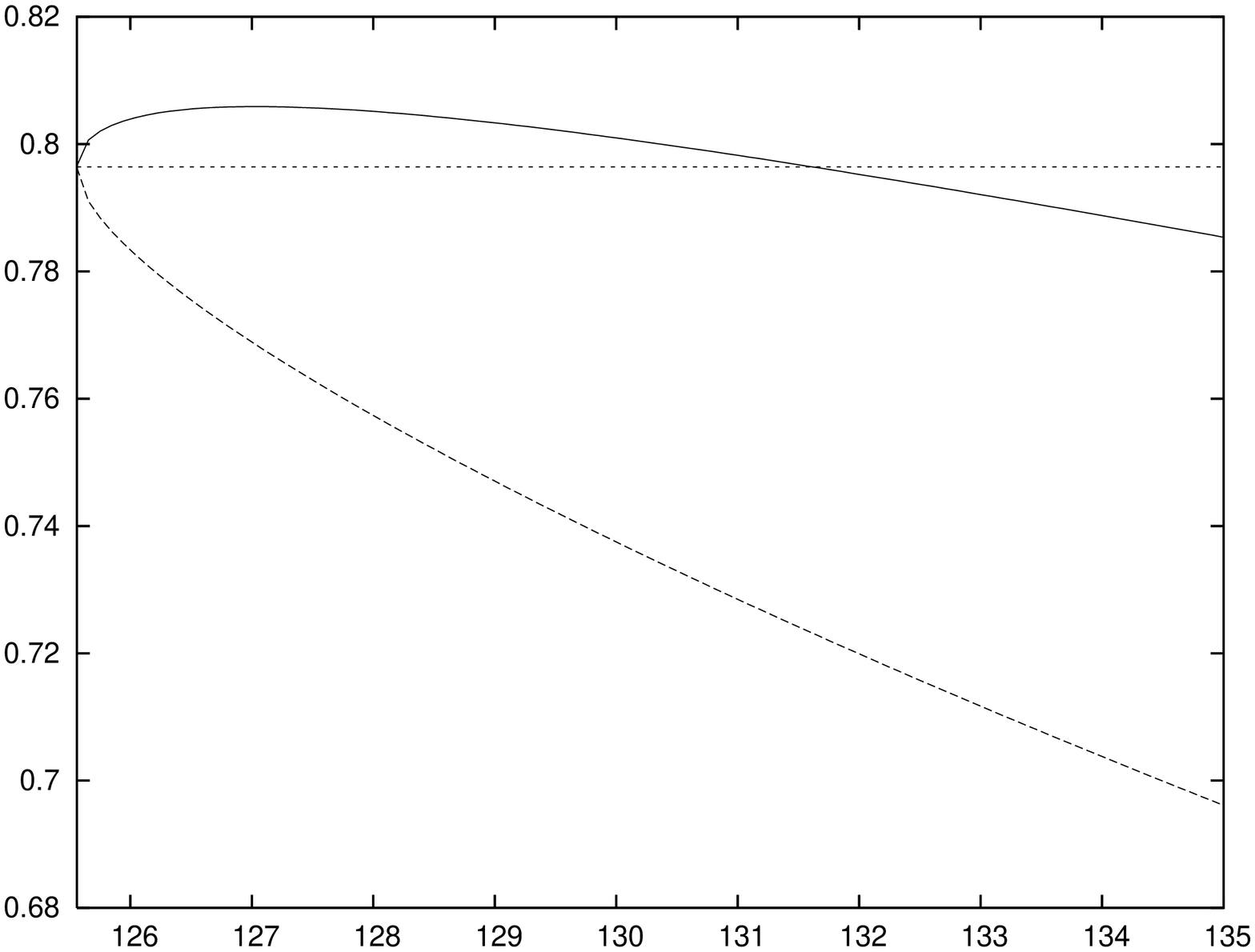}}
\end{picture}
\vspace{0.2cm}
\caption{Level-crossing in the case of a non-extremal RN-AdS black hole, 
with $L=10$,$M=10000$,$Q=100$, $m=10^{-3}$,$e=1$, $k=1$. The particle and the 
black hole have charges with the same sign. 
Black hole horizon occurs for $r\sim 125.56$.  
The straight line represents the value $e Q/r_+$. The upper potential is 
$\ezp$, the lower one is $\ezm$. Level-crossing occurs 
where $\ezp<e Q/r_+$. The figure on the right displays the potentials near $r=\rp$.}
\label{fig3}
\end{figure}

\begin{figure}[htbp!]
\begin{picture}(150,150)
\put(0,160){\includegraphics[height=.34\textheight,
                      width=.24\textheight,angle=-90]{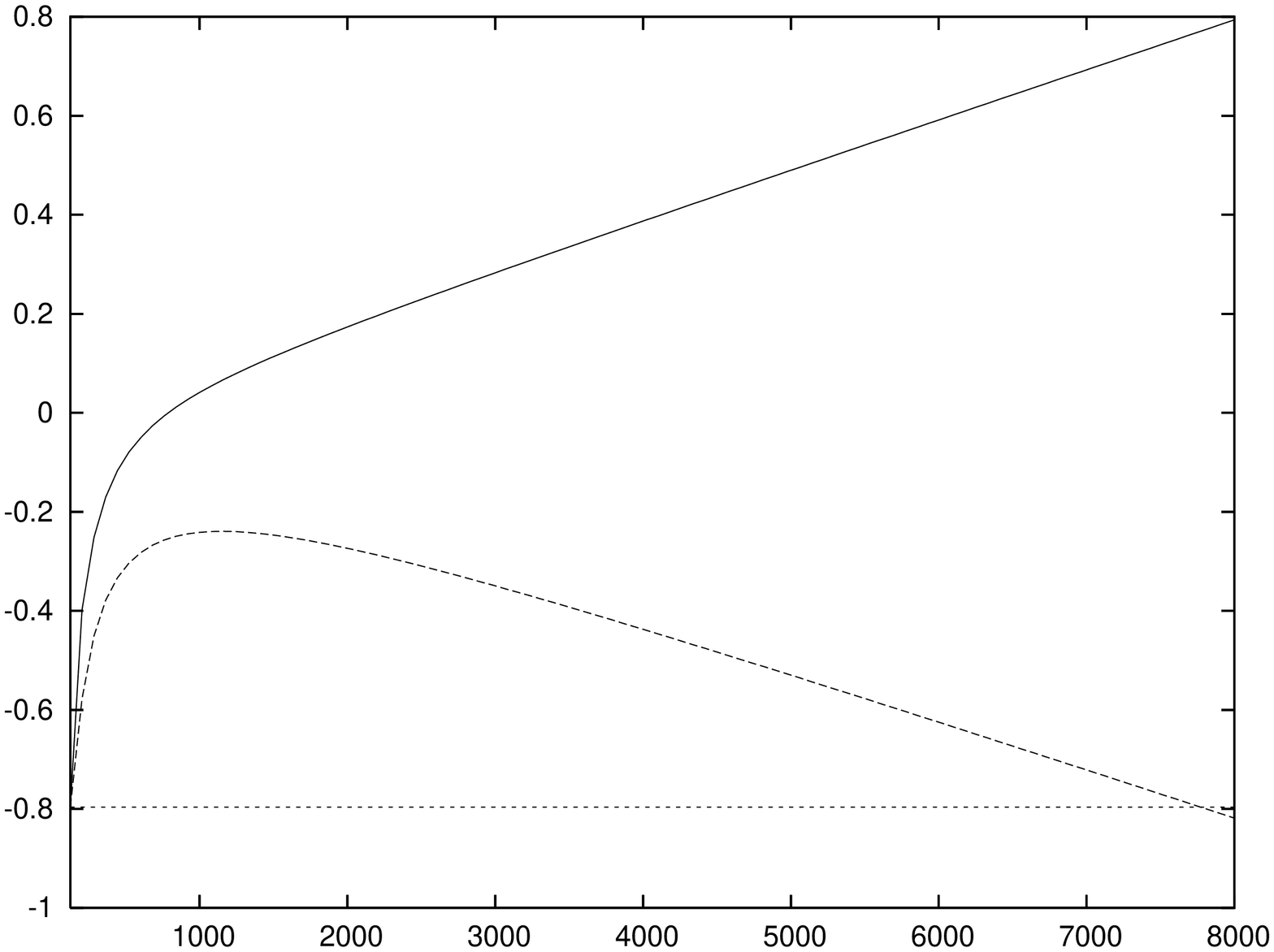}}
\put(230,160){\includegraphics[height=.30\textheight,
                      width=.24\textheight,angle=-90]{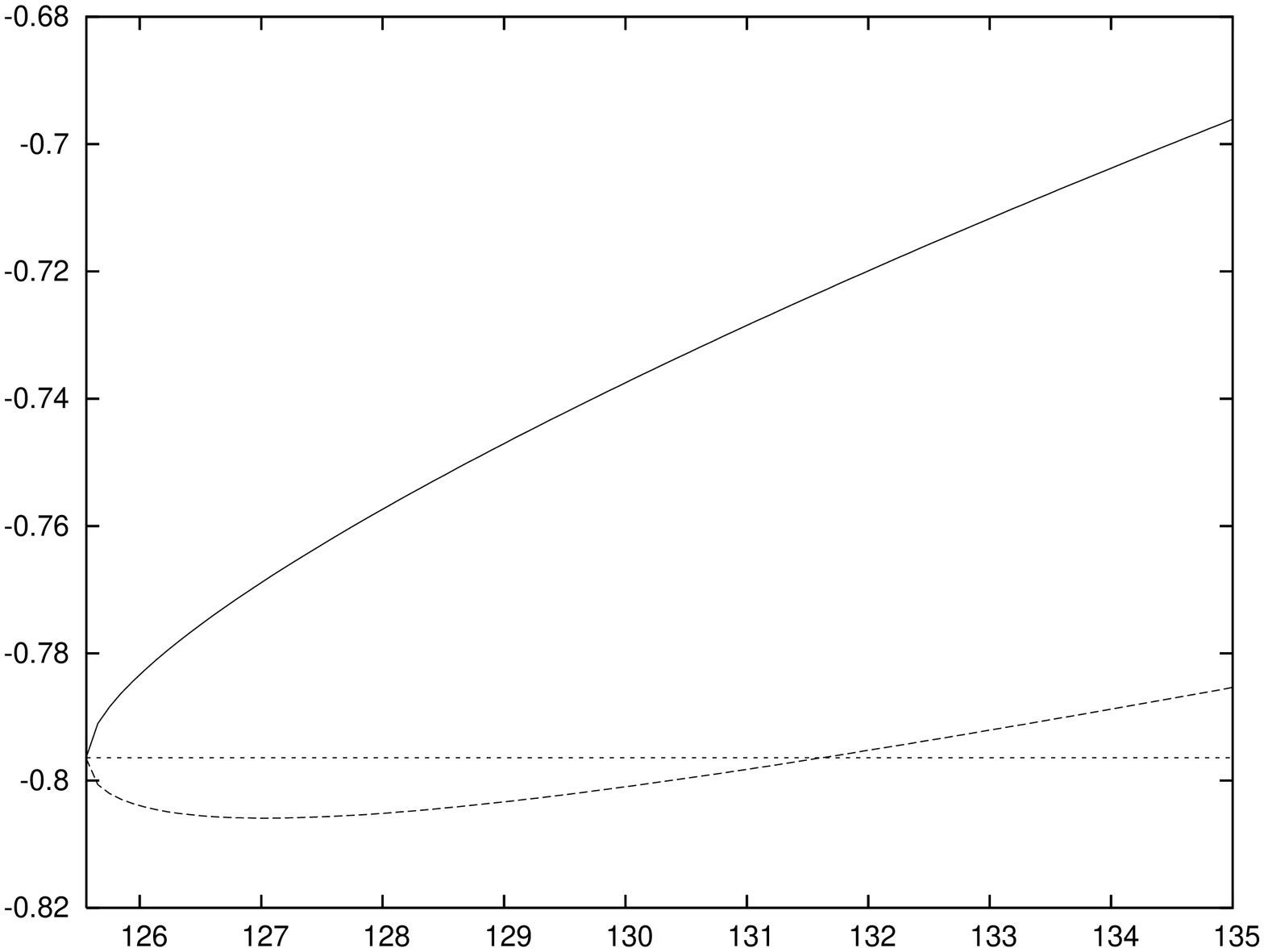}}
\end{picture}
\vspace{0.2cm}
\caption{Level-crossing in the case of a non-extremal RN-AdS black hole, 
with $L=10$,$M=10000$,$Q=100$, $m=10^{-3}$,$e=-1$, $k=1$. The particle and the 
black hole have charges with opposite sign. 
Black hole horizon occurs for $r\sim 125.56$.  
The straight line represents the value $e Q/r_+$. The upper potential is 
$\ezp$, the lower one is $\ezm$. Level-crossing occurs 
where $\ezp<e Q/r_+$. The figure on the right displays the potentials near $r=\rp$.}
\label{fig4}
\end{figure}

In the general case, it is difficult to find out analytically 
the overlap region between positive and negative energy states. 
It is still possible to find out level-crossing by means 
of graphical tools. We show below two examples. They concern 
the case of an extremal RN-AdS black hole and of a non-extremal 
one, respectively. In fig. \ref{fig1} the extremal case is displayed in the case that 
the particle and the black hole have charge with the same sign; for the same 
parameters, in fig. \ref{fig2} the opposite sign case is shown. Analogously, 
in fig. \ref{fig3} and fig. \ref{fig4} the non-extremal case is displayed.



\subsection{The meaning of $E_0^{\pm}(r)$ in the case of the Dirac Hamiltonian}
\label{meaning}

Let us consider the potential term in the Dirac Hamiltonian:
\[ V(r)=\left[
\begin{array}{cc}
p_{11}(r) &  p_{12}(r) \cr 
p_{21}(r)
& p_{22}(r)\end{array} \right]. 
\]
One can formally calculate the eigenvalues of the above matrix, which are 
found by solving
\beq
(p_{11}(r)-\lambda)(p_{22}(r)-\lambda)-p_{12}(r) p_{21}(r)=0;
\eeq
then, defining $S(r)\equiv \sqrt{(p_{11}(r)+p_{22}(r))^2-4 p_{11}(r) p_{22}(r)
+4 p_{12}(r) p_{21}(r)}$ one finds
\beq
\lambda^{\pm} (r)=\frac{1}{2}\left( 
p_{11}(r)+p_{22}(r)\pm S(r)\right). 
\eeq
One has
\beq
\lambda^{\pm} (r)=E_0^{\pm}(r),
\eeq
i.e., the potentials coincide with the eigenvalues of the matrix 
potential term in the Dirac Hamiltonian. 
Moreover, one has 
\beq
E_0^{\pm}(r)=\frac{eQ}{r}\pm \frac{1}{r^2} \sqrt{\Delta (\mu^2 r^2+k^2)}.
\eeq
Note again that the square of the classical angular momentum term is replaced by the 
square of the eigenvalues $k=\pm (j+1/2)$ for the quantum angular momentum. 
Then, in the case of a Dirac particle, the angular momentum contribution 
$k$ cannot vanish.\\
From the point of view of qualitative spectral analysis, the eigenvalues-potentials 
$\lambda^{\pm} (r)$ play a relevant role in the following sense. Let us consider for 
simplicity the reduced Dirac Hamiltonian in the case of flat spacetime; 
referring e.g. to theorems 16.5 and 16.6 of \cite{weidmann}, which concern the essential spectrum 
of Dirac systems, one finds that: in theorem 16.5 a fundamental role is played by the 
eigenvalues $\nu_{-}\leq \nu_+$ of the matrix $P_0\equiv \lim_{r\to \infty} P(r)$. It is 
evident that
\beq
\nu_{\pm}=\lim_{r\to \infty} \lambda^{\pm} (r).
\eeq 
In theorem 16.6 the norm $||P(t)-P_0||$ can be replaced with the norm of the 
difference 
$\left(
\begin{array}{cc}
\lambda^{+} (t) &  
0\cr
0  
& \lambda^{-} (t) 
\end{array} 
\right)-
U(t) P_0 U^{\dagger} (t)$, where the unitary matrix $U(t)$ diagonalizes $P(t)$ and 
where $\lim_{t\to \infty} U(t) P_0 U^{\dagger} (t)= \left(
\begin{array}{cc}
\nu_{+} &  
0\cr
0  
& \nu_{-} 
\end{array} 
\right)$
(this is quite obvious if one chooses to define the norm of any matrix $P$ as a norm 
in a suitable Hilbert space ${\mathcal H}$: $||P||\equiv \sup_{v\in {\mathcal H}} \frac{||P v||}{||v||}$, 
because in this case $||P||=||UPU^{\dagger}||$ for any unitary matrix $U$).\\  
The aforementioned results can be applied also to the black hole case. 
See e.g. \cite{belgio} and also \cite{yamada}.\\

Some physical considerations are in order. In presence of an external field (like e.g. 
an external electrostatic field and a static non-flat spacetime metric) the properties of the 
second-quantized theory cannot be related in a straightforward way to the spectral decomposition of 
the one-particle Hamiltonian $H$ into positive and negative energy states. This 
of course happens in the free case in flat spacetime, but, in presence of external fields, 
an intrinsic relevance of the 
classical external potentials in determining the structure of the quantum field theory 
associated with the given one-particle Hamiltonian has to be stressed. 
In the given case, the 
potentials $\lambda^+$ and $\lambda^-$ not only delimit the allowed regions of the classical 
motion, but also give physical insights about the characteristics of the quantum field theory one 
can build up by starting from the given $H$. On the one hand, their role is analogous to the 
one of the potential in non-relativistic quantum mechanics 
for a single particle: the potential defines the allowed regions of the classical motion, 
gives rises to quantum resonances and to quantum tunnelling phenomena and its behavior at 
infinity fixes also the limits of the essential spectrum (see e.g. theorems 
16 and 17, pp. 1448-1449 of \cite{dunford} for simple mathematical results which can be 
applied to one-dimensional quantum mechanical systems and in particular to the 
radial Schr\"odinger equation). On the other hand, there is a further requirement 
which has to be satisfied if a well-defined second quantization of the Dirac field 
has to be obtained. It is the requirement that positive and negative continuous energy states must be 
separated the ones from the others. It may happen that 
a phenomenon of level-crossing occurs, i.e. would-be negative energy states can overlap 
with positive energy ones. This lack of a clear separation between the positive and the negative 
energy solutions of the Dirac equation 
can be physically interpreted as the occurrence of an instability phenomenon 
which leads to pair creation \cite{damo}. In particular, defining as in Ref. \cite{damo} 
a complete set of positive and negative `in' modes $p_i^{in}$ and $n_i^{in}$ and 
a complete set of `out' modes $p_i^{out}$ and $n_i^{out}$, with the channel for the 
transition $n_k^{in}\to p_i^{out}$ a non-vanishing transition amplitude
\beq
T_{ik}=(p_i^{out},n_k^{in})
\eeq
can be associated. The mean number of pairs which are created is \cite{damo}
\beq
<N>=\sum_{{\mathrm{ channels}}\; i,k} |T_{ik}|^2.
\eeq
This should be compared with the transmission 
coefficient in the Klein paradox situation, where the transmission coefficient is
\beq
|T|^2= \frac{|{\mathrm{ transmitted\; flux}}|}{|{\mathrm{ incident\; flux}}|}.
\eeq
The absence of overlapping between the ranges of
$\lambda^+$ and $\lambda^-$ avoids the occurrence of level-crossing.  
When an overlap occurs, the possibility that a 
positron state can emerge in a suitable range as a positive energy scattering state 
is at the root of the Klein paradox and also of the pair-creation process. Cf. 
also \cite{thaller}.  
Semi-quantitative estimates for this phenomenon can be given e.g. in the WKB 
approximation. In the case analyzed herein, 
numerical estimates show that the presence of a non-vanishing $k$ makes  
more difficult to find a level-crossing (which, in the scalar particle case 
occurs more easily for $l=0$); nevertheless it is still possible to 
find out an overlap which is not limited to $r=\rp$:
\beq
\lambda^+ ((r_+,\infty))\cap \lambda^- ((r_+,\infty))\supset \lambda^+ (r_+)
\eeq

Some further consideration has to be addressed to the problem of the choice of the 
quantum state playing the role of vacuum. In the extremal black hole case, the so-called 
Boulware vacuum, which corresponds to the positive and negative frequencies arising from 
the Hamiltonian, is allowed and then the previous analysis is enough. In the non-extremal 
case, one has to take into account the fact that the regular quantum state corresponds to 
the so-called Hartle-Hawking state, which is the thermal state at the Hawking temperature, 
whereas the Boulware vacuum is singular on the horizon. Suitable analyticity requirements 
for the fields on the extended manifold lead to the thermal state \cite{gibbons}.  
Notwithstanding, the thermal state 
can be obtained by ``heating up'' the Boulware one at least in the Schwarzschild case, where  
the construction can be implemented rigorously for the scalar field \cite{kay}. For the 
Dirac field see \cite{melnyk}.  
In presence of electric charge, like in the present case, there is 
a further difficulty which is associated with the presence of level-crossing (it has 
manifestly the characteristics of the Klein paradox in the Reissner-Nordstr\"om case) 
which signals the presence of a further source of particle creation which is different 
from the one associated with the Hawking effect.

\subsection{A picture for a discharge}

\begin{figure}[htbp!]
\begin{picture}(150,160)
\put(80,160){\includegraphics[height=.44\textheight,
                      width=.34\textheight,angle=-90]{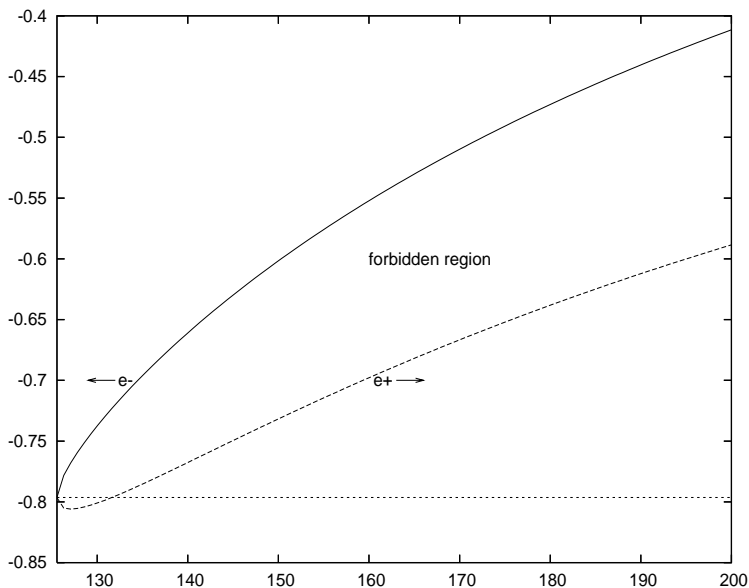}}
\end{picture}
\vspace{2.1cm}
\caption{Pair creation in the case of a non-extremal RN-AdS black hole, 
with $L=10$,$M=10000$,$Q=100$, $m=10^{-3}$,$e=-1$, $k=1$. The particle and the 
black hole have charges with opposite sign. 
Black hole horizon occurs for $r\sim 125.56$.  
The straight line represents the value $e Q/r_+$. The upper potential is 
$\ezp$, the lower one is $\ezm$.} 
\label{bhpair}
\end{figure}

According to the classical potentials described above, the negative energy state, 
for definiteness, the positron with suitable energy can tunnel beyond the barrier 
and reach the external region, i.e. in the classically allowed region  
where an observer can measure it and also determine 
that, by charge conservation, the charge of the black hole is diminished. 
This transition, in a WKB approximation, is exponentially suppressed. 
Near infinity, the positron meets a classical turning point where it is 
reflected back towards the barrier. 
The possibility of a further passage of the barrier is 
exponentially suppressed, and then the whole process of emission-re-absorption 
of the positron by the black hole is suppressed with respect to 
the simple emission process. As a consequence, the phenomenon of pair-creation 
and emission of positrons by a positively charged black hole 
is allowed also in the RN-AdS black hole manifold. 
See also fig. \ref{bhpair}. 

The emitted positrons 
are mostly confined in the classically allowed region, and they classically 
behave as bound states with a minimal and a maximal distance from the black hole. 
They do not correspond to stable quantum mechanical bound states, indeed 
there is no discrete spectrum for the Hamiltonian, and, as a consequence, they 
cannot be identified with isolated eigenvalues of the Hamiltonian. They could at most 
correspond to resonances (cf. also \cite{damo}). This picture holds 
as far as back-reaction effects by the emitted positrons are negligible. 
In a WKB approximation, one finds \cite{damo}
\beq
|T|^2 = \exp (-2 \int_{\mathrm{barrier}} dz \sqrt{Z}).
\eeq

\subsection{Discussion}

There is a problem in the definition of a sensible notion of particle in 
the RN-AdS background. This kind of difficulty occurs also in the case 
of the pure AdS manifold. 
The above discussion about the presence of level-crossing in the RN-AdS case 
relies on the fact that the classical potentials 
$E_0^{\pm}(r)$ still maintain their heuristic meaning in the RN-AdS case as in 
the standard asymptotically flat RN case. The difference is that one 
cannot define asymptotic scattering states for the Hamiltonian as 
$r\to \infty$ in the RN-AdS case as well as in the AdS one. 
Note that, in the latter case, one has 
\beq
E_0^{\pm}(r)=\pm \frac{1}{r^2} \sqrt{\left(\frac{r^4}{L^2}+r^2\right) (\mu^2 r^2+k^2)}.
\eeq
No level-crossing occurs, as expected. The behavior as $r\to \infty$ 
of $E_0^{\pm}(r)$ is the same as for the  RN-AdS background, at the 
leading order.

\subsection{A note on the case $\mu L < \frac{1}{2}$}

In the case where essential self-adjointness is missing, one can in the scalar 
field case impose reflective boundary conditions (Dirichlet) and also Neumann 
boundary conditions as in the pure AdS case. In the case of the Dirac field, 
one can e.g. impose the MIT bag boundary conditions which play the same role 
for the Dirac field as the Dirichlet boundary conditions for the scalar field; if 
$\Psi$ is the four spinor and $\vec{n}$ is the (outer) normal to the boundary $\pa \Omega$ 
of the domain $\Omega$, the MIT boundary condition amount to 
\beq
i \vec{\gamma}\cdot \vec{n} \Psi|_{\pa \Omega} = \Psi|_{\pa \Omega}.
\eeq
It is easy to show that, in the spherosymmetrical case, these boundary condition amount to 
\beq
{g_1}|_{\pa \Omega} = {g_2}|_{\pa \Omega}.
\eeq
By using the same coordinates as in sec. \ref{sect-infinity}, one finds the 
condition  
\beq
{g_1}(x=0) = {g_2} (x=0).
\eeq
One can also introduce other local boundary conditions, see e.g. \cite{bachelot} 
for the case of pure AdS. We don't take into account any longer this topic herein.

\section{Conclusions}

We have studied the self-adjointness properties of the Dirac Hamiltonian in the 
RN-AdS black hole background. We have found that the horizon does not require any 
boundary condition, whereas $r=+\infty$ may behave as a boundary also in the Dirac case. 
Nevertheless, if the bound $|\mu L|\geq 1/2$ is implemented, then essential self-adjointness 
is ensured. In the latter case, we have taken into account the spectral properties 
of the Hamiltonian, and we have found that the essential spectrum is the whole real line. 
This means that the presence of an horizon also in this case is sufficient for ensuring 
that no discrete spectrum (which is by definition made of isolated eigenvalues of finite 
multiplicity) occurs. Moreover, a study of the potentials $\lambda^{\pm}$ 
which arise in a semi-classical approximation enabled us to verify that a level crossing 
between states with energy $E>\lambda^+$, which correspond to classically allowed 
states, and states with energy $E<\lambda^{-}$, which correspond to the classical 
counterpart of the Dirac sea states, can occur.  
This overlapping, as well-known from previous literature on this topic, indicates the 
possibility to find out a phenomenon of pair-creation by means of tunneling from 
the Dirac sea to the classically allowed states. WKB approximation can be implemented in 
order to describe the tunneling. As a consequence, a way to describe the discharge of the 
black hole by means of a quantum effect which shares a different origin with respect to 
the Hawking effect (and, in fact, it occurs also in the case of extremal black holes) 
has been shown to be available also in the case of Dirac fields on RN-AdS black hole 
backgrounds.

\ack
F.B. wishes to thank Professors Vittorio Gorini and Ugo Moschella (Universit\`a dell'Insubria, 
Como) for useful discussions 
at the early stages of this work.

\appendix

\section{$M=0=Q$ case: pure AdS}

It is useful to introduce AIS (Avis-Isham-Storey) coordinates:
\beq
ds^2= L^2\; \frac{1}{\cos^2 (\rho)}\; 
\left[ - dt^2+d\rho^2 +\sin^2 (\rho)\; d\Omega^2 \right],
\label{ais}
\eeq
where $0\leq \rho<\pi/2$ and $t\in \RR$ (CAdS).  
One finds 
\[ H_{red}=\left[
\begin{array}{cc}
\frac{1}{\cos \rho}\; \mu &  
-\frac{1}{L}\;  \partial_{\rho} +k\; \frac{1}{L}\; 
\frac{1}{\sin \rho}\cr
\frac{1}{L}\; \partial_{\rho} +k\; \frac{1}{L}\; 
\frac{1}{\sin \rho}  
& -\frac{1}{\cos \rho}\;  \mu 
\end{array} 
\right] 
\]
It is easy to see that, in the limit as $\rho\to 0^+$ the LPC is 
verified, that is, one does not need to impose a boundary condition at 
$\rho=0$ ($r=0$ in the original coordinates). In order to study the 
limit as $\rho \to (\pi/2)^-$ it is useful to introduce a new coordinate 
$x\equiv (\pi/2)-\rho$. Then 
\[ H_{red}=\left[
\begin{array}{cc}
\frac{1}{\sin x}\; \mu &  
\frac{1}{L}\;  \partial_{x} +k\; \frac{1}{L}\; 
\frac{1}{\cos x}\cr
-\frac{1}{L}\; \partial_{x} +k\; \frac{1}{L}\; 
\frac{1}{\cos x}  
& -\frac{1}{\sin x}\;  \mu 
\end{array} 
\right] 
\]
One gets the 
following system of first order equations for the eigenvalue equation:
\[ x \vec{g}'=\left[
\begin{array}{cc}
k \frac{x}{\cos x} &
- \frac{x}{\sin x}\; \mu L-\lambda L x \cr 
- \frac{x}{\sin x}\; \mu L+\lambda L x &
-k \frac{x}{\cos x}
\end{array} \right] \vec{g}\equiv A(x)  \vec{g};
\]
the prime indicates the derivative w.r.t. x. The matrix $A(x)$ is regular as $x\to 0$ 
and one has 
\[ \lim_{x\to 0} A(x)\equiv A_0 =
\left[
\begin{array}{cc}
0  &
-\mu L\cr 
-\mu L & 0 
\end{array} \right].   
\]
The same considerations as in subsection \ref{sect-infinity} apply. 
Restoring the physical constants, one finds the following condition for 
essential self-adjointness (see also \cite{bachelot}):
\beq
\mu c^2\geq \frac{1}{2}\; (\hbar c)\; \frac{1}{L}\equiv \mu_{bound}.
\eeq

\section{LPC at the horizon}

For the non-extremal case, we define
\beq
z=r-\rp, 
\eeq
then the eigenvalue equation for $H_{red}$ can be re-written as follows:
\[ z \vec{g}'=\left[
\begin{array}{cc}
-\frac{k \sqrt{z}}{(\rp+z) \sqrt{h(z)}} &
 \frac{\sqrt{z}}{\sqrt{h(z)}}\; \mu -(\frac{e\; Q}{\rp+z} -\lambda)\; \frac{1}{h(z)} \cr 
 \frac{\sqrt{z}}{\sqrt{h(z)}}\; \mu +(\frac{e\; Q}{\rp+z} -\lambda)\; \frac{1}{h(z)} &
\frac{k \sqrt{z}}{(\rp+z) \sqrt{h(z)}}
\end{array} \right] \vec{g}; 
\]
the prime stays for the derivative with respect to $z$ and $h(z)$ is 
defined by 
\beq
h(z) = \frac{f(z)}{z}.
\eeq
The matrix on the left of the above system is regular at $z=0$ and its limit as $z\to 0$ 
is given by the constant matrix 
\[ A_0=\left[
\begin{array}{cc}
0  &
-\frac{1}{h(0)} (\Phi_+ - \lambda)\cr
 \frac{1}{h(0)} (\Phi_+ - \lambda) & 0
\end{array} \right],   
\]
where $\Phi_+$ is the electrostatic potential at the horizon. The eigenvalues of the 
above matrix are
\beq
\epsilon_{\pm}=\pm i \frac{1}{h(0)} (\Phi_+ - \lambda).
\eeq
As in the previous subsection, one can find the asymptotic behavior of the solution 
as $z\to 0$. The leading order is given by $z^{\epsilon_{\pm}}$. It is easy to see that 
no solution is integrable near the horizon. Let us consider 
\beq
\lambda= \lambda_{R}+i \lambda_{I};
\label{comlam}
\eeq 
then, one finds
\beq
\int_0^{\delta} \frac{dz}{z h(z)} |z^{\epsilon_{\pm}}|^2=
\int_0^{\delta} \frac{dz}{z h(z)} z^{\pm 2 \frac{\lambda_{I}}{h(0)}};
\eeq
for any choice of $\lambda\in \CC$, there is a solution which is divergent near $z=0$. 
As a consequence, the LPC holds at the horizon.\\ 
The extremal case is more difficult because an irregular singularity appears for $r=\rp$. 
The change of variable
\beq
r=\rp+1/x,
\eeq
with $x\in (0,\infty)$ allows us to find a system in the form $\vec{g}'= A \vec{g}$, with 
\[A=\left[
\begin{array}{cc}
 k \frac{\sqrt{f}}{r B} & \frac{1}{B} ( -\mu\sqrt{f}+\frac{e\; Q}{r} -\lambda)\cr
 -\frac{1}{B} ( \mu\sqrt{f}+\frac{e\; Q}{r} -\lambda) & - k \frac{\sqrt{f}}{r B} 
\end{array} \right];   
\]
the prime stays for the derivative with respect to $x$ and 
\beq
B=\frac{1}{L^2} \frac{1}{r^2} (r^2+2\rp r+3 \rp^2+L^2).
\eeq
One has
\[
\lim_{x\to \infty} B \equiv B_0 = \left[
\begin{array}{cc}
 0 & \frac{1}{B(0)} (\frac{e\; Q}{\rp} -\lambda)\cr
 -\frac{1}{B(0)} ( \frac{e\; Q}{\rp} -\lambda) & 0 
\end{array} \right],   
\]
whose eigenvalues are
\beq
\eta_{\pm}=\pm i \frac{1}{B(0)} (\frac{e\; Q}{\rp} -\lambda).
\eeq
There exists a nonsingular matrix $T$ such that $A^{diag}\equiv T A T^{-1}$ tends to a diagonal matrix 
$A_0^{diag}$ with entries $\eta_{\pm}$ for $x\to \infty$. 
$A^{diag}$ admits a power series expansion in some neighborhood of $x=\infty$:
\beq
A^{diag}=\sum_{k=0}^{\infty} A_k^{diag} \frac{1}{x^k}.
\eeq 
Applying the general theory \cite{coddington} to the present case, a formal solution matrix $\Psi$ can be found 
in the form
\beq
\Psi=P x^R \exp(Q x),
\eeq
where $P$ is a formal power series 
\beq
P=\sum_{k=0}^{\infty} P_k \frac{1}{x^k}, \qquad \det (P_0)\not = 0,
\eeq
$Q$ coincides with $A_0^{diag}$ and $R$ is a diagonal constant matrix whose entries coincide with 
the diagonal entries of $A_1^{diag}$. 
The diagonal entries of $A_1^{diag}$ are given by
\beq
\zeta_{\pm}=\pm i \left(L^2 e Q +\lambda - \frac{e Q}{\rp}\right)
\left(2 \rp L^2 -\frac{4 \rp^3 L^2}{6\rp^2+L^2}\right) \frac{1}{6\rp^2+L^2}. 
\eeq
By using (\ref{comlam}) one finds that the condition of $L^2$-integrability of the leading order 
solution at $x=\infty$ cannot be satisfied by both the solutions, because 
\beq
\int_c^{\infty} \frac{dx}{B} x^{\pm 4 \rp L^2 \frac{4\rp^2+L^2}{6\rp^2+L^2}\lambda_{I}} 
\exp (\pm \frac{2}{B(0)} \lambda_{I} x) 
\eeq 
is divergent for one of them for any choice of $\lambda\in \CC$. Then the LPC case occurs.\\

\end{document}